\documentclass{eas}
\usepackage{graphicx}
\usepackage{amssymb}
\usepackage{amsmath}

\begin{document}

\title{Nonlinear tidal flows in short-period planets}
\author{Adrian J. Barker}
\address{Department of Applied Mathematics, School of Mathematics, University of Leeds, Leeds, LS2 9JT, UK} 
\secondaddress{Department of Applied Mathematics and Theoretical Physics, University of Cambridge, Centre for Mathematical Sciences, Wilberforce Road, Cambridge CB3 0WA, UK}

\begin{abstract}
I discuss two related nonlinear mechanisms of tidal dissipation that require finite tidal deformations for their operation: the elliptical instability and the precessional instability. Both are likely to be important for the tidal evolution of short-period extrasolar planets. The elliptical instability is a fluid instability of elliptical streamlines, such as in tidally deformed non-synchronously rotating or non-circularly orbiting planets. I summarise the results of local and global simulations that indicate this mechanism to be important for tidal spin synchronisation, planetary spin-orbit alignment and orbital circularisation for the shortest period hot Jupiters. The precessional instability is a fluid instability that occurs in planets undergoing axial precession, such as those with spin-orbit misalignments (non-zero obliquities). I summarise the outcome of local MHD simulations designed to study the turbulent damping of axial precession, which suggest this mechanism to be important in driving tidal evolution of the spin-orbit angle for hot Jupiters. Avenues for future work are also discussed.
\end{abstract}
\maketitle

\section{Introduction}
\begin{figure}
\begin{center}
\includegraphics[width=0.605\textwidth]{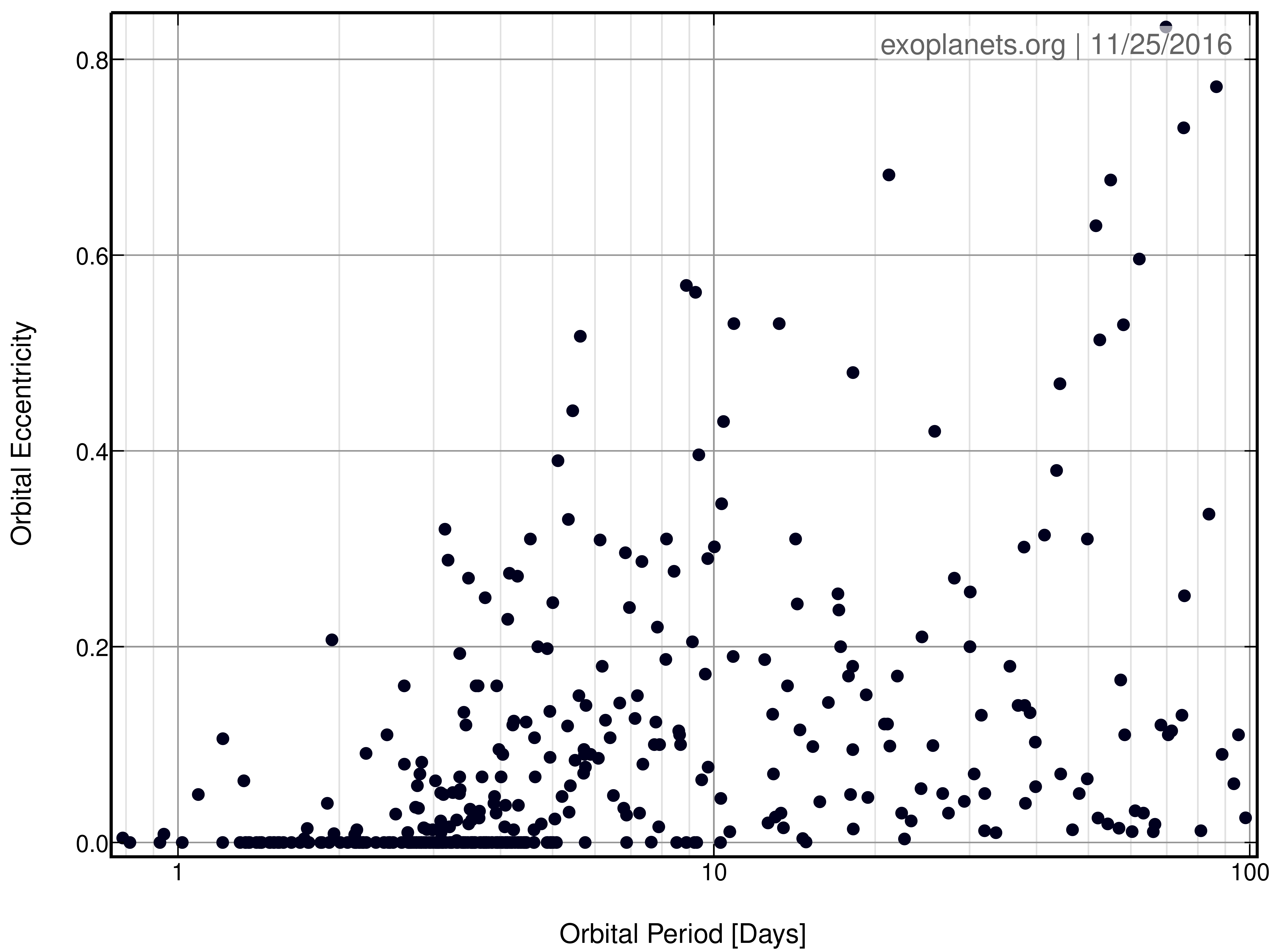}
\end{center}
\caption{Eccentricity distribution of Jupiter-mass extrasolar planets (with $M\geq 0.3 M_J$ and $P<100$ d). Planets with $P>10$ d have a wide range of eccentricities, whereas the population with $P<10$ d has much lower eccentricities and displays a strong preference for circular orbits. This provides strong evidence of the important role of tidal dissipation in shaping planetary orbits.}
\label{evsP}
\end{figure}

Tidal interactions between short-period planets and their host stars are thought to play an important role in the evolution of the planetary orbit as well as the stellar and planetary spins (e.g.~Zahn \cite{Z13}; Mathis et al. \cite{M13}; Ogilvie \cite{O14}). The clearest evidence of tidal evolution in extrasolar planetary systems is the eccentricity distribution of the approximately Jupiter-mass planets (here taken to mean masses $M\geq 0.3 M_J$), which is shown in Fig.~\ref{evsP}. Planets with $P>10$ d have a wide range of eccentricities, whereas the population with $P<10$ d has much lower eccentricities and displays a strong preference for circular orbits. Tidal dissipation inside the planet (and perhaps partly also the star) is thought to be responsible for this dichotomy, since it tends to damp the orbital eccentricities of the shortest-period planets (it may also have played a role in their formation if they underwent a period of ``high eccentricity migration" to attain such periods e.g. Wu \& Lithwick \cite{WL11}; Naoz et al. \cite{N11}; Anderson et al. \cite{ASL16}). The timescale for this evolution depends on the mechanisms of tidal dissipation, which are only partly understood theoretically. Here, I will focus on nonlinear tidal mechanisms that require finite amplitude tidal deformations. This is partly because nonlinear effects are the least well understood, but more importantly it is because their consideration is likely to be essential for understanding the tidal evolution of short-period planets.

The (dimensionless) tidal deformations of short-period planets can be estimated by (the height of the tide is approximately $\epsilon_T R_p$)
\begin{eqnarray}
\epsilon_T = \frac{m_{\star}}{m_p}\left(\frac{R_p}{a}\right)^3 \approx 0.01 \left(\frac{P}{1\,\mathrm{d}}\right)^{-2},
\end{eqnarray}
where $m_\star$ and $m_p$ are the stellar and planetary masses, $R_p$ is the planetary radius, $a$ is the orbital semi-major axis, and $P$ is the orbital period (taking $m_p=M_J$, $m_\star=M_{\odot}$ and $R_p=R_J$ on the right hand side). The most extreme current example is WASP-19 b (Hebb et al. \cite{H09}), with its $0.78$ d orbital period, giving $\epsilon_T\sim 0.05$. This is no longer a small parameter, indicating that nonlinear effects could be important even for large-scale tidal flows in such a body. This can be compared with the tides in Jupiter and Saturn due to their most massive satellites ($\epsilon_T\sim 10^{-7}$), where nonlinear effects may be much less important for the largest-scale tidal flows (though they could still be important in damping tidally-excited short-wavelength waves).

In this paper, I will discuss two related nonlinear tidal mechanisms: the elliptical instability and the precessional instability. The former occurs in fluids with elliptical streamlines (see also the related paper by Favier \cite{F17}), such as in tidally deformed planets, and the latter occurs in fluid bodies that undergo axial precession, such as planets with misaligned spins and orbits (nonzero obliquities). Both are parametric instabilities driven by the periodic time-dependence of fluid properties around a streamline, leading to the excitation of inertial waves (restored by the Coriolis force) in the planetary interior. And both mechanisms are likely to be important for tidal evolution of short-period planets. 

\section{Elliptical instability}

\begin{figure}
\begin{center}
\raisebox{0.3in}{\includegraphics[trim=0cm 0cm 0cm 0cm, clip=true,width=0.4\textwidth]{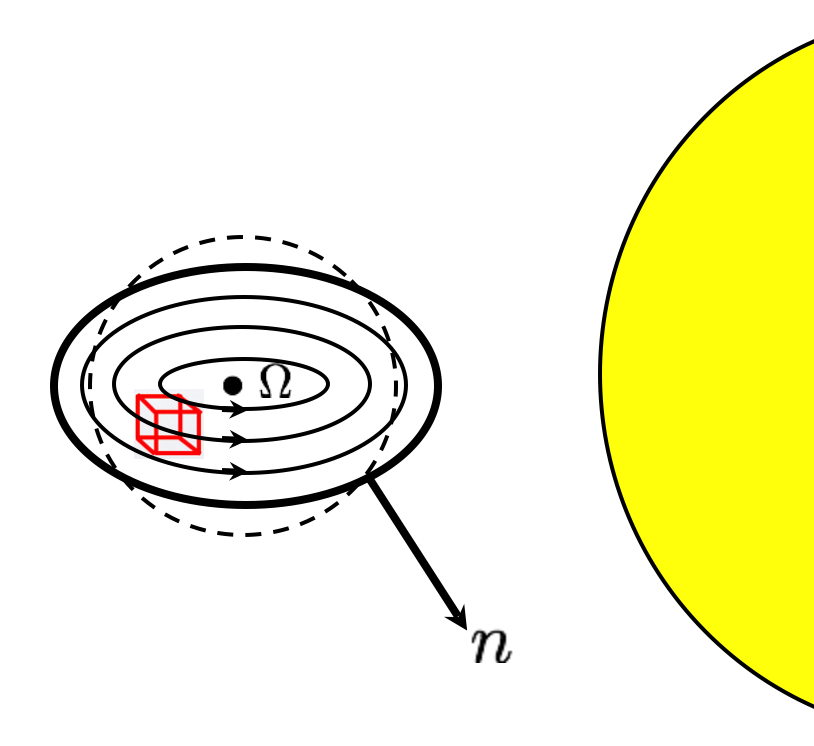}} \hspace{0.05in}
\includegraphics[trim=11cm 0cm 14cm 1cm, clip=true,width=0.46\textwidth]{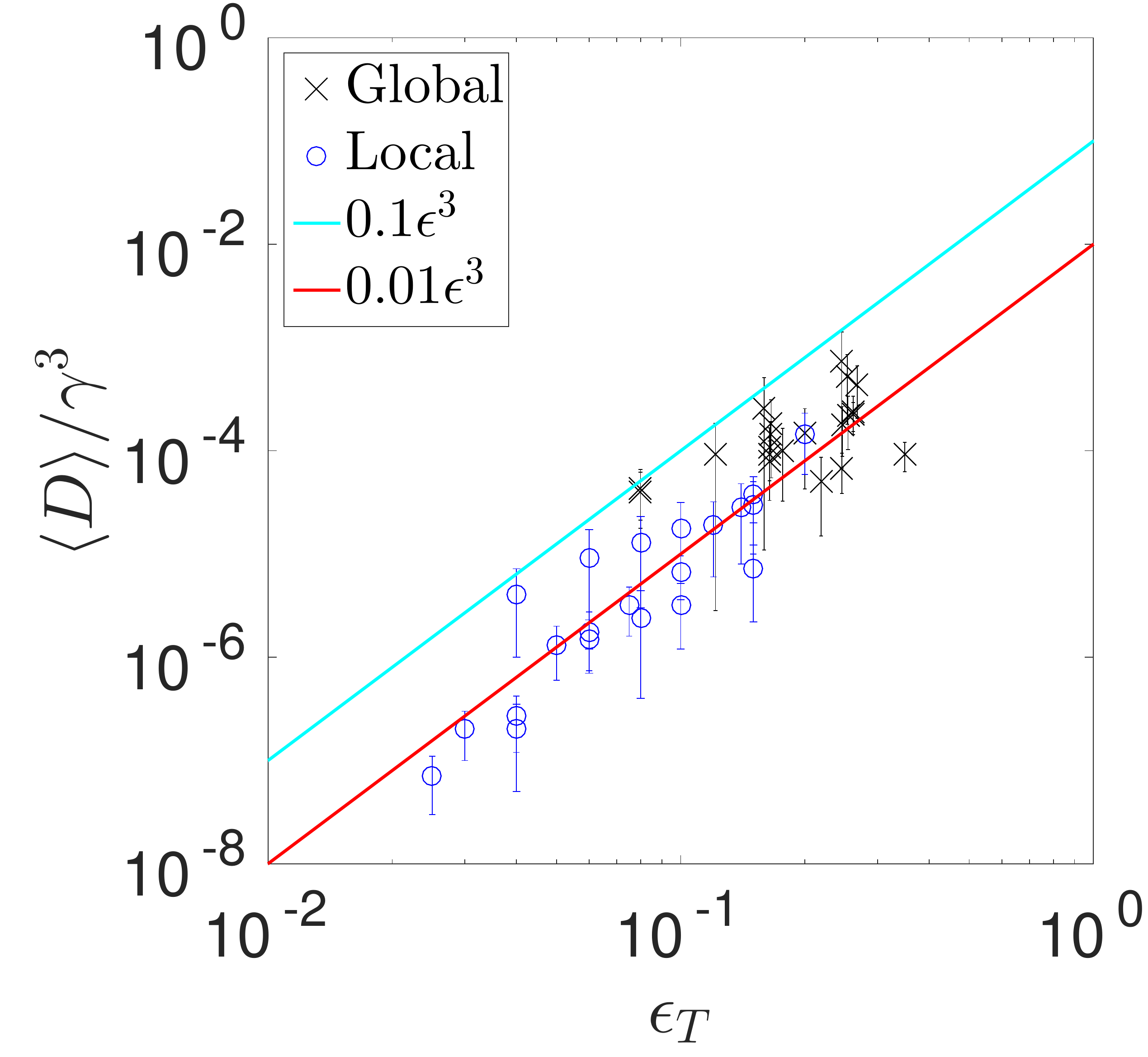}
\end{center}
\caption{Left: Illustration of global elliptical flow inside the planet viewed from above (the rotation axis $\Omega$ is pointing towards us), also indicating the local model considered by Barker \& Lithwick (\cite{BL13,BL14}). Right: Results of local Cartesian (MHD) and global ellipsoidal (hydrodynamic) simulations of the elliptical instability, showing that the turbulent (volume and time-averaged) dissipation is consistent with a simple cubic scaling with $\epsilon_T\gamma$.}
\label{EI}
\end{figure}

The elliptical instability is a fluid instability of elliptical streamlines, such as the large-scale non-wave-like tidal flow in a planet that is non-synchronously rotating or has an eccentric orbit (see the left panel of Fig.~\ref{EI} for illustration). The simplest case for illustration is a planet on a circular orbit but that is non-synchronously rotating (with an aligned or anti-aligned spin axis). In the frame rotating at the rate $\Omega$ about the spin axis, the elliptical deformation has frequency $2\gamma$, where $\gamma=n-\Omega$. Inertial waves exist with frequencies $|\omega | < 2|\Omega|$. If a pair of these waves (with subscripts 1 \& 2) has frequencies such that $|\omega_1\pm \omega_2|\approx 2 |\gamma|$, then the superposition of one of these waves with the deformation can excite the other wave, and vice versa, leading to instability. Consideration of the spatial structure of the waves leads to the additional requirement that the azimuthal wavenumbers and harmonic degrees satisfy $|m_1\pm m_2|=2$ (since the deformation has $m=2$) and $\ell_1=\ell_2$. The maximum growth rate (which typically occurs when $|\omega_1|\approx |\omega_2|\approx \gamma$) is (e.g.~Kerswell \cite{K02})
\begin{eqnarray}
\sigma \sim \epsilon_T\gamma f(n,\gamma),
\end{eqnarray}
where $f(n, \gamma)$ is a dimensionless function of $n$ and $\gamma$. In the limit $\epsilon_T\ll 1$, instability occurs if $-1 \leq \frac{n}{\Omega} \leq 3$, but is also possible outside of this range (particularly for anti-aligned spins with $\frac{n}{\Omega}\lesssim-1$, if $\epsilon_T$ is sufficiently large, due to the finite resonant widths e.g.~Barker et al. \cite{BBO16}).

The instability leads to turbulence in the planetary interior that produces enhanced tidal dissipation. In hydrodynamical simulations (Barker \cite{B16a}), the instability generates differential rotation in the planetary interior in the form of zonal flows (or columnar vortices in a local Cartesian model, at least if $\epsilon_T\lesssim 0.15$; Barker \& Lithwick \cite{BL13}), which control the saturation of the instability, leading to cyclic, predator-prey-like dynamics (where zonal flows act as the predator and waves the prey) with intermittent dissipation. Zonal flows have also been observed in experiments of the elliptical instability (e.g.~Le Bars et al. \cite{LB15}; Favier et al. \cite{F15}). In the presence of weak magnetic fields (Barker \& Lithwick \cite{BL14}), or an alternative frictional mechanism that damps zonal flows (e.g.~Le Reun et al. \cite{LFBL17}), the instability leads to sustained turbulence with significantly enhanced dissipation over that of the laminar tidal flow.

The (volume and time-averaged) turbulent dissipation rate $D$ resulting from sustained instability can be simply estimated. Nonlinearities saturate growth of the unstable waves if $u \lambda^{-1}\sim \sigma$, where $u$ is a typical velocity and $\lambda$ is the wavelength of the dominant waves. Kinetic energy is dissipated at the rate $D\sim \sigma u^2 m_p \sim \epsilon_T^3\gamma^3\lambda^{2}m_p$, where we have taken $f(n,\gamma)=1$ for simplicity. For numerically accessible $\epsilon_T \sim 0.01-0.3$ (including the relevant regime for the shortest-period planets such as WASP-19 b), both local and global simulations are consistent with this simple argument, as we show in Fig.~\ref{EI}. We find $D\approx \chi_E \epsilon_T^3 \gamma^3R_p^{2} m_p$, where $\chi_E\approx 0.1$ provides an approximate upper bound on $D$ (though $\chi_E$ does vary to some extent with $n$, as does $\sigma$). It remains to be seen whether this scaling would persist for much smaller $\epsilon_T$, when the instability is more likely to excite very short-wavelength waves ($\lambda\ll R_p$).  Also, these arguments neglect the inhibiting effects of zonal flows that lead to intermittent rather than sustained turbulence. Hence, this scaling is best thought of as an approximate upper bound on the dissipation (due to bulk turbulence) if we want to extrapolate to smaller $\epsilon_T$.

In the frequency range where the instability operates, the resulting turbulent dissipation would drive tidal synchronisation (and evolution of the planetary spin-orbit angle $\psi$) on a timescale
\begin{eqnarray}
\tau_\Omega\approx \tau_\psi \approx 1\,\mathrm{Gyr}\, \left(\frac{0.1}{\chi_E}\right)\left(\frac{P}{14.7 \,\mathrm{d}}\right)^{6}\left(\frac{P_\mathrm{rot}}{1 \,\mathrm{d}}\right),
\end{eqnarray}
where $P_\mathrm{rot}$ is the planetary rotation period, and circularisation on a timescale
\begin{eqnarray}
\tau_e \approx 1\,\mathrm{Gyr}\, \left(\frac{0.1}{\chi_E}\right)\left(\frac{P}{2.8 \,\mathrm{d}}\right)^{\frac{25}{3}},
\end{eqnarray}
where we have taken the planet to have Jupiter's mass and radius (and radius of gyration) and the star to be Sun-like for both estimates\footnote{Note that these are much more strongly period dependent than a linear tidal mechanism or by a naive constant $Q$ (or constant lag-time) model.} (Barker \cite{B16a}). Hence, this instability could be important for tidal evolution of the shortest-period hot Jupiters, and we would predict orbital circularisation for $P\lesssim 3$ d, and spin synchronisation and alignment for $P\lesssim 15$ d, due to this mechanism acting in isolation. However, additional mechanisms appear to be required to explain tidal circularisation for $3\,\mathrm{d} \lesssim P\lesssim 10\,\mathrm{d}$ (Fig.~\ref{evsP}), such as tidal dissipation of linearly-excited inertial waves (and their linear or nonlinear damping) in planets with a core (e.g. Ogilvie \& Lin \cite{O04}), or dissipation in the solid core itself (Remus et al. \cite{R12}; Storch \& Lai \cite{S14}).

\section{Precessional instability}

\begin{figure}
\begin{center}
\includegraphics[trim=12cm 0cm 13.5cm 1cm, clip=true,width=0.47\textwidth]{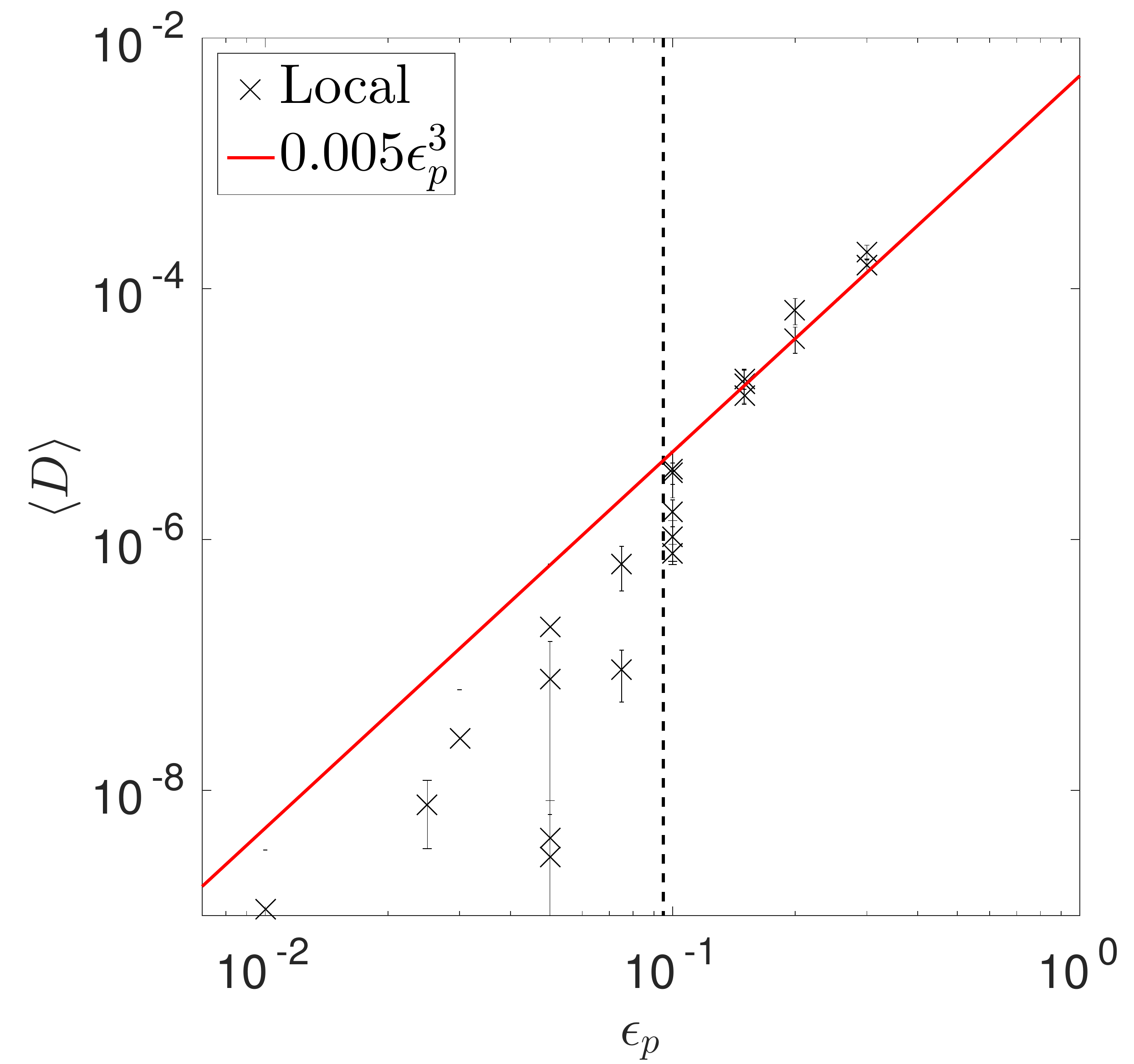}
\includegraphics[trim=12cm 0cm 13.5cm 1cm, clip=true,width=0.47\textwidth]{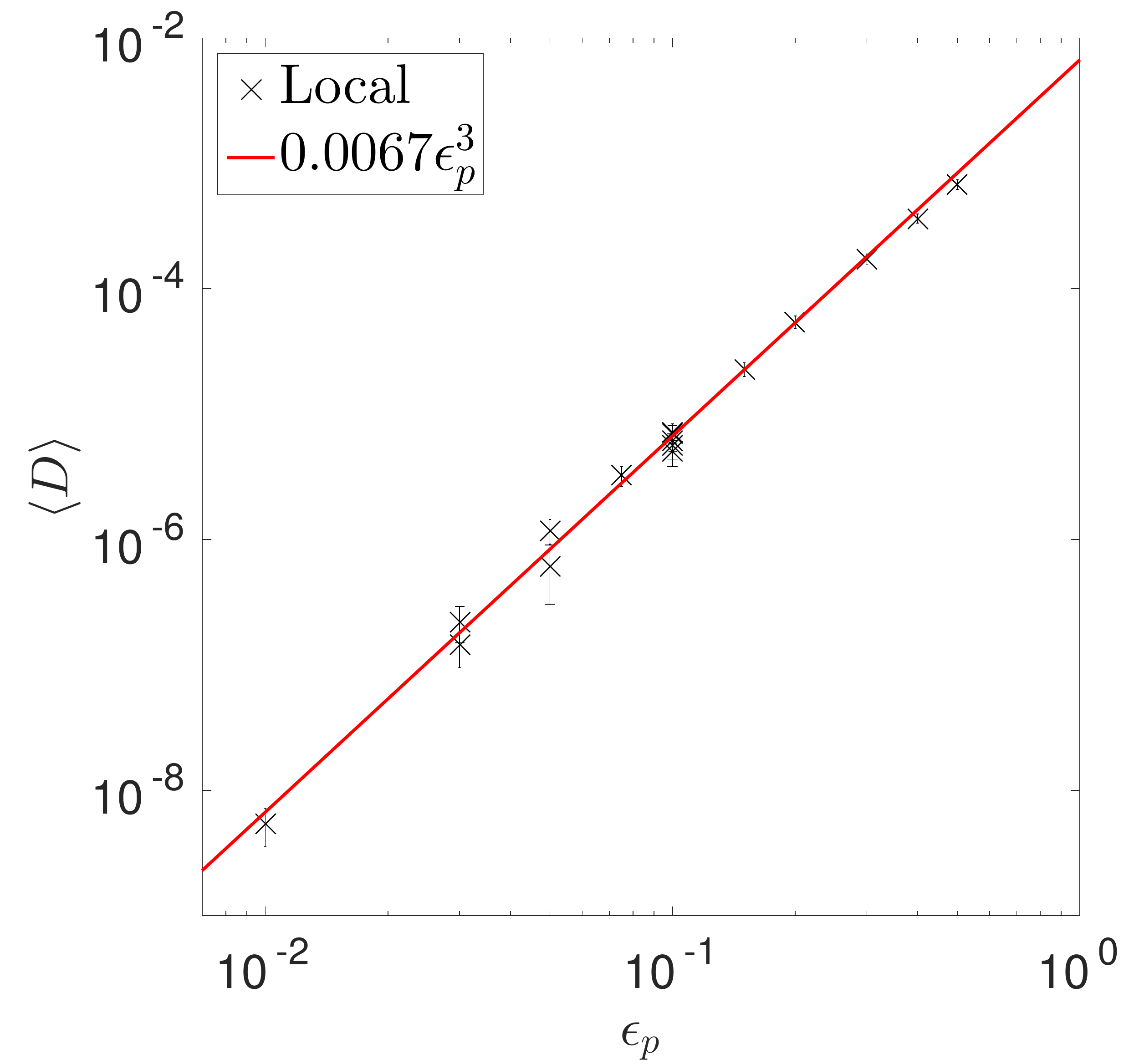}
\end{center}
\caption{Turbulent (volume and time-averaged) dissipation from local Cartesian hydrodynamic (left), and MHD (right) simulations with an initially weak magnetic field as a function of the dimensionless precession rate $\epsilon_p$. There is a regime transition in the hydrodynamic simulations, where columnar vortices inhibit sustained instability for $\epsilon_p\lesssim 0.1$. The MHD simulations are consistent with a simple cubic scaling with the precession rate over the simulated range, and appear to be approximately independent of the diffusivities.}
\label{PI}
\end{figure}

The spin axis of a rotationally deformed planet is forced to precess about its orbital angular momentum vector, due to the tidal gravity of its host star, if these directions are misaligned (if the spin-orbit angle $\psi\ne 0,90^{\circ},180^{\circ}$). A typical period of axial precession is
\begin{eqnarray}
P_p = \frac{2\pi}{\Omega_p}\approx 1.3 \,\mathrm{yr} \left(\frac{P}{10\,\mathrm{d}}\right)^2 \left(\frac{P_{\mathrm{rot}}}{10\,\mathrm{hr}}\right),
\end{eqnarray}
where $\Omega_p\equiv \epsilon_p \Omega\propto \epsilon_T\Omega$ is the precession rate ($\epsilon_p$ is a dimensionless version), for a Jupiter-mass and radius planet orbiting a Sun-like star. This precession is potentially observable for some transiting planets through transit depth variations (e.g.~Carter \& Winn \cite{CW10}), which would provide important constraints on the oblateness and interior structure of the planet. However, axial precession would only occur in single planet systems (and be potentially observable) if $\psi$ can remain significantly misaligned. Tidal evolution is expected to ultimately lead to alignment ($\psi\rightarrow 0^{\circ}$). One mechanism that could play a role in driving this tidal evolution is the precessional instability, which will now be introduced.

Axial precession induces internal flows inside the planet that are time-periodic, with frequency $|\Omega|$ in the frame rotating with the planet. Pairs of inertial waves can be driven unstable if $|\omega_1\pm\omega_2|\sim |\Omega|$, and the fastest growing modes typically have $|\omega_1|\sim|\omega_2|\sim |\Omega|/2$, with a maximum growth rate (Kerswell \cite{K93})
\begin{eqnarray}
\sigma \sim \Omega_p.
\end{eqnarray}

The nonlinear evolution of this instability in a local Cartesian model behaves just like the elliptical instability (Barker \& Lithwick \cite{BL13}, Barker \cite{B16b}). For small $\epsilon_p$, the purely hydrodynamical instability generates columnar vortices which inhibit sustained instability and lead to cyclic, predator-prey-like dynamics, with intermittent turbulent dissipation. The addition of a weak magnetic field eliminates these vortices and leads to sustained turbulence. Global simulations are currently in progress, where we expect the instability to generate zonal flows, and for these to subsequently control the dissipation (this was demonstrated already in the specific case of an initially anti-aligned spin, once it had evolved away from pure anti-alignment, in Barker \cite{B16a}, where the elliptical instability was also present).

The turbulent dissipation resulting from this instability can be estimated in the same way as for the elliptical instability outlined above. The local MHD (and also, approximately, the ``bursty" hydrodynamic) simulations are consistent with $D\approx\chi_p \Omega_p^3R_p^{2}m_p$ (with turbulent velocities $u\sim \Omega_p R_p$), with $\chi_p\approx 0.01$, over the range $\epsilon_p\sim 0.01-0.5$, as we show in Fig.~\ref{PI}. This range is somewhat larger than the realistic values of $\epsilon_p$, requiring us to extrapolate the scaling to probe the astrophysical regime. It remains to be seen whether this scaling would persist for much smaller $\epsilon_p$, so it is best thought of as an approximate upper bound on the bulk turbulent dissipation when $\epsilon_p\ll 1$.

The consequences of this turbulent dissipation are tidal evolution of the spin-orbit angle $\psi$ for short-period planets, which evolves due to this mechanism acting in isolation according to (Lai \cite{L12}, Ogilvie \cite{O14})
\begin{eqnarray}
\frac{d\psi }{d t}=-\frac{\sin\psi\cos^2\psi}{\tau_{\psi}}\left(\cos\psi + \alpha\right),
\end{eqnarray}
where $\alpha$ is the ratio of planetary spin to orbital angular momentum and
\begin{eqnarray}
\tau_\psi \approx 1 \,\mathrm{Gyr} \left(\frac{0.01}{\chi_p}\right)\left(\frac{P}{18 \,\mathrm{d}}\right)^6\left(\frac{P_\mathrm{rot}}{10\,\mathrm{hr}}\right)
\end{eqnarray}
is the timescale for $\psi$ to evolve towards alignment ($\psi=0^{\circ}$) if $|\psi|$ is small (Barker \cite{B16b}), for a Jupiter-mass and radius planet around the current Sun. This is also the timescale for $\psi$ to evolve towards anti-alignment ($\psi=180^{\circ}$), if $|\psi-180^{\circ}|$ is small. The corresponding timescale if $\psi\sim 90^{\circ}$ is larger by a factor $\alpha^{-1}\gtrsim 10^3$, indicating that most planets with nearly perpendicular spins will not undergo appreciable evolution by this mechanism acting in isolation. However, other tidal mechanisms will also contribute, and in general we expect the ultimate tidal evolution to be towards alignment. Nevertheless, this crude estimate suggests that the precessional instability could be important for tidal evolution of the spin-orbit angle for the shortest-period hot (and warm) Jupiters with $P\lesssim 18$ d.

\section{Discussion}
I have discussed two related nonlinear mechanisms of tidal dissipation that require finite-amplitude tidal deformations for their operation, which could both be important for the spin-orbit evolution of short-period extrasolar planets. 

The elliptical instability occurs in tidally deformed fluid bodies. Local and global simulations suggest that it could drive tidal spin synchronisation (and spin-orbit alignment) for hot (and warm) Jupiters with\footnote{The lower limit to these ranges is based on a modified scaling of $D$ with $\epsilon_T$ that takes into account that $\lambda \ll R_p$ when $\epsilon_T\ll 1$ (Barker \cite{B16a}), which is also consistent with simulations for accessible values of $\epsilon_T\gtrsim 0.01$.} $P\lesssim 10-15$ d, and tidal circularisation for $P\lesssim 2-3$ d. Other mechanisms are probably required to explain the small eccentricities of planets with $3 \,\mathrm{d}\lesssim P\lesssim 10\,\mathrm{d}$, shown in Fig.~\ref{evsP}. 

The precessional instability occurs in axially precessing fluid bodies, such as in rotating planets with spin-orbit misalignments. Simulations using a local model suggest that it can drive tidal evolution of the spin-orbit angle (towards alignment or anti-alignment, depending on the initial value) for planets with $P\lesssim 10-18$ d (Barker \cite{B16b}). Global simulations of this instability are currently in progress.

The models that have been adopted so far to study these instabilities are extremely idealised, mainly restricted to homogeneous fluid bodies. Real planets and stars are certainly not homogeneous, but exploring such models constituted a necessary first step. Future work is required to simulate the instability in more realistic planetary models (i.e.~with a realistic density and entropy stratification), including the presence of an inner core (e.g.~by extending Favier et al.~\cite{FBBO14} to adopt more realistic boundary conditions and consider larger tidal amplitudes), as well as the influence of magnetic fields in global models. The effect of zonal flows on the saturation of these instabilities in the astrophysically relevant regime of small deformations (or precession rates), as well as vanishingly small viscosities should be considered, as should the coexistence of both types of instability, and these instabilities with turbulent convection. Clearly, much further work is required to understand tidal dissipation in planets and stars, and for us to be able to explain the astrophysical observations.

\section{Acknowledgements}
AJB was funded by a Leverhulme Trust Early Career Fellowship.

\end{document}